\documentclass[a4paper,12pt]{article}
\usepackage[english]{babel}
\usepackage{amsfonts,amssymb,amsmath}

\begin{document}

 \centerline{\bf\Large Renormalized asymptotic solutions} 
\vskip 1mm 
 \centerline{\bf\Large of the Burgers equation} 
\vskip 1mm 
 \centerline{\bf\Large and the Korteweg--de Vries equation}

 \vskip 5mm 
 \centerline{\bf S.V.~Zakharov} 

 \vskip 5mm 

\begin{center}
Institute of Mathematics and Mechanics,\\
Ural Branch of the Russian Academy of Sciences,\\
16, S.Kovalevskaja street, 620990, Ekaterinburg, Russia

\vskip 2mm 

E-mail: {\tt svz@imm.uran.ru}
\end{center}

\vskip 5mm 

\textbf{Abstract.}
The Cauchy problem for the Burgers equation
and the Korteweg--de Vries equation is considered.
Uniform renormalized asymptotic solutions
are constructed in cases of a large initial gradient
and a perturbed initial weak discontinuity.


 
\vskip 5mm 

\section{Problem with a large initial gradient} 

A simplest model of the motion of continuum,
which takes into account nonlinear effects and dissipation,
 is the equation of nonlinear diffusion 
\begin{equation}\label{eqw}  
 \frac{\partial u}{\partial t} + 
 u\frac{\partial u}{\partial x} = 
 \varepsilon \frac{\partial^2 u}{\partial x^2},
\qquad \varepsilon>0,
\end{equation}
for the first time presented by J.~Burgers~\cite{bu}. 
 This equation  is used in studying
 the evolution of a wide class of physical systems and probabilistic
processes, for example, acoustic waves in fluid and gas~\cite{Gbw}. 

Let us consider the Cauchy problem
\begin{equation}\label{qpic}
 u(x,0,\varepsilon,\rho) = \Lambda ( {x}{\rho}^{-1}),
 \quad t = 0, \quad x\in\mathbb{R},  \quad \rho>0,
\end{equation}
where $\rho$ is another small parameter. 
The initial function~$\Lambda$ is smooth
and it has finite limits
$$\Lambda^\pm_0=\lim\limits_{s\to\pm\infty}\Lambda(s). $$

We see that the origin ($x=0$ and $t=0$)
must be a singular point 
in the problem under consideration. 
It is known that the behavior of solutions of differential equations
with a  small parameter at a higher derivative
in a neighborhood of a singular point
sometimes becomes self-similar. 
Then it is effective to analyze the solution
using the renormalization group method~\cite{bk2}. 
The success of applying renormalization is
confirmed by examples from hydrodynamics~\cite{gv},
mechanics~\cite{teodor}, 
and asymptotic analysis of solutions
of nonlinear parabolic equations~\cite{bkl}.

In paper~\cite{zvm}, for the solution
of a more general quasi-linear parabolic equation 
$$
 \frac{\partial u}{\partial t} + 
 \frac{\partial \varphi(u)}{\partial  x} = 
 \varepsilon \frac{\partial^2 u}{\partial  x^2},
 \quad t\geqslant 0, 
$$
with condition $(\ref{qpic})$
the following asymptotic formula is obtained: 
\begin{equation}\label{ab} 
 {u}(x,t,\varepsilon,\rho)= 
 \frac{1}{\Lambda^+_0 - \Lambda^-_0} 
 \int\limits_{-\infty}^{\infty} 
 \Gamma\left( \frac{x-\rho s}{\varepsilon}, \frac{t}{\varepsilon}\right) 
 \Lambda'(s)\, ds 
 +O\left(\left(\frac{\rho}{\varepsilon}\right)^{1/4}\right)
  \end{equation}
 as $\varepsilon\to 0$ and $\rho/\varepsilon\to 0$ 
 uniformly in the strip $\{ (x,t) :
x\in\mathbb{R},\ 0 \leqslant t\leqslant T\}$, 
 where $\Gamma$ is the solution of the limit problem 
 in the inner variables $\eta = x/\varepsilon$, 
 $\theta = t/\varepsilon$ 
$$ 
 \frac{\partial \Gamma}{\partial \theta} + 
 \frac{\partial \varphi(\Gamma)}{\partial \eta} - \frac{\partial^2 \Gamma}{\partial \eta^2} = 0, 
 \quad 
 \Gamma(\eta,0) = 
 \begin{cases} 
 \Lambda^-_0, & \eta<0, \\ 
 \Lambda^+_0, & \eta > 0. 
 \end{cases} 
 $$ 
The procedure of obtaining expression (\ref{ab})
can be also seen below in details on the example of the KdV equation.
The difference consists in the scales of the inner variables.
For the solution of the Burgers equation (\ref{eqw})
 as $\varepsilon\to 0$ and $\rho/\varepsilon\to 0$ 
 in the strip $\{ (x,t) : x\in\mathbb{R},\ 0 \leqslant
t\leqslant T\}$ 
 there holds the asymptotic formula $$ 
 {u}(x,t,\varepsilon,\rho)= 
 \int\limits_{-\infty}^{\infty} 
 \frac{\Lambda'(s)}{\Lambda^+_0-\Lambda^-_0} 
 \left[ \Lambda^+_0 \exp \left( 
 \frac{t(\Lambda^+_0)^2-2\Lambda^+_0 x}{4\varepsilon} 
 +\frac{\Lambda^+_0 \rho s}{2\varepsilon}\right) 
 \mathrm{erfc}\left( 
 \frac{\Lambda^+_0 t -x +\rho s}{2\sqrt{\varepsilon
t}} 
 \right)+ 
 \right. 
 $$ 
 $$ 
 +\left.\Lambda^-_0 \exp \left( 
 \frac{t(\Lambda^-_0)^2-2\Lambda^-_0 x}{4\varepsilon} 
 +\frac{\Lambda^-_0 \rho s}{2\varepsilon}\right) 
 \mathrm{erfc}\left( 
 \frac{x - \Lambda^-_0 t -\rho s}{2\sqrt{\varepsilon
t}} 
 \right) 
 \right] \times 
 $$ 
 $$ 
 \times \left[\exp\left( 
 \frac{t(\Lambda^+_0)^2-2\Lambda^+_0 x}{4\varepsilon} 
 +\frac{\Lambda^+_0 \rho s}{2\varepsilon}\right) 
 \mathrm{erfc}\left( 
 \frac{\Lambda^+_0 t -x +\rho s}{2\sqrt{\varepsilon
t}} 
 \right) 
 +\right. 
 $$ 
 $$ 
 +\left. \exp \left( 
 \frac{t(\Lambda^-_0)^2-2\Lambda^-_0 x}{4\varepsilon} 
 +\frac{\Lambda^-_0 \rho s}{2\varepsilon}\right) 
 \mathrm{erfc}\left( 
 \frac{x - \Lambda^-_0 t -\rho s}{2\sqrt{\varepsilon
t}} 
 \right) 
 \right]^{-1} ds 
 +O\left(\mu^{1/4}\right). 
 $$

 \bigskip 
\setcounter{equation}{0}
\section{Renormalized weak discontinuity}

In the present paper, we give an asymptotic solution       
of the Burgers equation 
\begin{equation}\label{bn}  
 \frac{\partial u}{\partial t} + 
 u\frac{\partial u}{\partial x} = 
 \frac{\partial^2 u}{\partial x^2}
\end{equation}
with the initial condition
\begin{equation}\label{icw}
 u(x,0) = \varepsilon \Lambda\left( \frac{x}{\varepsilon}\right), 
 \quad t=0,  \quad x\in\mathbb{R}, 
\end{equation}
where
$\Lambda(s) \to -s$ as $s\to -\infty$
and $\Lambda(s) \to 0$ as $s\to +\infty$.

By analogy with (\ref{ab}) we postulate the expression
$$
R(x,t,\varepsilon) =
\int\limits_{-\infty}^{\infty} \Lambda''(s)
u_0(x-\varepsilon s,t) \, ds,
$$
where $u_0 = -2 \Psi_x/\Psi$ is the solution
of the limit problem with a weak discontinuity~\cite{vm4},
$$
 \frac{\partial u_0}{\partial t} + 
 u_0\frac{\partial u_0}{\partial x} = 
 \frac{\partial^2 u_0}{\partial x^2},
\qquad
 u_0(x,0) = 
 \begin{cases} 
-x, & x<0, \\ 
\phantom{-}0, & x \geqslant 0. 
 \end{cases} 
$$
 $\Psi$ is the solution
of the heat equation $\Psi_t = \Psi_{xx}$.
Thus, we obtain a renormalized asymptotic solution
in the form
$$
R(x,t,\varepsilon) =
-\frac{2}{\varepsilon}
\int\limits_{-\infty}^{\infty} \Lambda'''(s)
\ln\Psi(x-\varepsilon s,t) \, ds.
$$

\vskip 3mm
\noindent
\textbf{Theorem~1.}
{\it Let $\Lambda(s)$ be a thrice differentiable function and let
\begin{equation}\label{l1m}
\frac{d\Lambda(s)}{ds} = -1 + O(|s|^{-3}),
\qquad s\to -\infty,
\end{equation}
\begin{equation}\label{l1p}
\frac{d\Lambda(s)}{ds} = O(s^{-3}),
\qquad s\to +\infty,
\end{equation}
\begin{equation}\label{l2}
\frac{d^2\Lambda(s)}{ds^2} = O(s^{-4}),
\qquad s\to \infty.
\end{equation}
Then the function
$$
R(x,t,\varepsilon) =
-\frac{2}{\varepsilon}
\int\limits_{-\infty}^{\infty}
\frac{d^3\Lambda(s)}{ds^3}
\ln\Psi(x-\varepsilon s,t) \, ds
$$
is an asymptotic solution of the Burgers equation $(\ref{bn}),$
where 
 $$ 
 \Psi(x,t) = \int\limits_{-\infty}^{0} 
 \exp\left( -\frac{(x-\sigma)^2}{4t} 
 +\frac{\sigma^2}{4}\right) d\sigma 
 +\int\limits_{0}^{\infty} 
 \exp\left( -\frac{(x-\sigma)^2}{4t}\right)
d\sigma.
 $$ 
}
\vskip 2mm
\noindent
\textbf{Proof.} Substituting $u=R(x,t,\varepsilon)$
into (\ref{bn}), we find
$$
R_t + RR_x - R_{xx} =
\frac{4}{\varepsilon^2}
\int\limits_{-\infty}^{\infty} \Lambda'''(s)
\frac{\Psi_x(x-\varepsilon s,t)}{\Psi(x-\varepsilon s,t)} \, ds
\int\limits_{-\infty}^{\infty} \Lambda'''(s)
\ln\Psi(x-\varepsilon s,t) \, ds -
$$
$$
- \frac{2}{\varepsilon}
\int\limits_{-\infty}^{\infty} \Lambda'''(s)
\frac{\Psi^2_x(x-\varepsilon s,t)}{\Psi^2(x-\varepsilon s,t)} \, ds.
$$
Let us esimate the integrals on the right-hand side.
From (\ref{l1m})--(\ref{l2}) it follows that
$$
\int\limits_{-1/\sqrt{\varepsilon}}^{1/\sqrt{\varepsilon}}
\Lambda'''(s) ds = O(\varepsilon^2),
\qquad
\int\limits_{-1/\sqrt{\varepsilon}}^{1/\sqrt{\varepsilon}}
s \Lambda'''(s) ds = -1+ O(\varepsilon^{3/2}).
$$
Then we obtain
$$
\int\limits_{-\infty}^{\infty} \Lambda'''(s)
\ln\Psi(x-\varepsilon s,t) \, ds
= \int\limits_{-1/\sqrt{\varepsilon}}^{1/\sqrt{\varepsilon}}
\Lambda'''(s) \left[
\ln\Psi(x,t) - \varepsilon s \frac{\Psi_x(x,t)}{\Psi(x,t)}
\right] ds + O(\varepsilon^2) =
$$
$$
= \varepsilon  \frac{\Psi_x(x,t)}{\Psi(x,t)} + O(\varepsilon^2),
$$
$$
\int\limits_{-\infty}^{\infty} \Lambda'''(s)
\frac{\Psi_x(x-\varepsilon s,t)}{\Psi(x-\varepsilon s,t)} \, ds
= - \varepsilon
\left( \frac{\Psi_{xx}}{\Psi}- \frac{\Psi^2_x}{\Psi^2}\right) 
\int\limits_{-1/\sqrt{\varepsilon}}^{1/\sqrt{\varepsilon}}
s \Lambda'''(s) ds + O(\varepsilon^{2}) =
$$
$$
= \varepsilon
\left( \frac{\Psi_{xx}}{\Psi}- \frac{\Psi^2_x}{\Psi^2}\right) 
 + O(\varepsilon^{2}),
$$
$$
\int\limits_{-\infty}^{\infty} \Lambda'''(s)
\frac{\Psi^2_x(x-\varepsilon s,t)}{\Psi^2(x-\varepsilon s,t)} \, ds
= - 2\varepsilon
\left( \frac{\Psi_x\Psi_{xx}}{\Psi^2}- \frac{\Psi^3_x}{\Psi^3}\right) 
\int\limits_{-1/\sqrt{\varepsilon}}^{1/\sqrt{\varepsilon}}
s \Lambda'''(s) ds + O(\varepsilon^{2})  =
$$
$$
=2 \varepsilon
\left( \frac{\Psi_x\Psi_{xx}}{\Psi^2}- \frac{\Psi^3_x}{\Psi^3}\right) 
 + O(\varepsilon^{2}).
$$
Consequently, we conclude that
$$
R_t + RR_x - R_{xx} = O(\varepsilon).
$$

Since $R(x,0,\varepsilon) = \varepsilon \Lambda(x/\varepsilon)$,
by the continuity of the inverse operator,
we can also state that $R(x,t,\varepsilon)$ is
an asymptotic approximation of the exact solution
of the Cauchy problem (\ref{bn})--(\ref{icw}). 

 \bigskip 
\setcounter{equation}{0}
\section{Korteweg--de Vries equation} 
 
 Studying properties of solutions 
 of the Korteweg--de Vries equation 
 \begin{equation}\label{eq} 
 \frac{\partial u}{\partial t} + 
 u\frac{\partial u}{\partial x} + 
 \varepsilon \frac{\partial^3 u}{\partial x^3} = 0, 
 \quad t\geqslant 0, \quad \varepsilon>0, 
 \end{equation} 
 is of indisputable interest 
 for describing nonlinear wave phenomena~\cite{Gbw}. 
 Open questions concerning the behavior of
solutions are still the subject of attention 
 for modern researches~\cite{gst}. 
 Some mathematical results about the solution of
the problem in various cases can be found
in~\cite{kf} and~\cite{fam}.
 
Consider the Cauchy problem for (\ref{eq}) with the initial
condition 
 \begin{equation} \label{ic}\phantom{\frac{1}{1}} 
 u(x,0,\varepsilon,\rho) = \Lambda ( {x}{\rho}^{-1}), 
 \quad t=0, \quad x\in\mathbb{R}, \quad \rho>0. 
 \end{equation} 
 
 Here, we assume that the initial function $\Lambda$ is
bounded and possesses the first derivative,
which sufficiently fast tends to zero at infinity. 
 It is required to find an asymptotic
approximation 
 of the solution $u(x,t,\varepsilon,\rho)$ 
 to problem~(\ref{eq})--(\ref{ic}) 
 in small parameters $\varepsilon$ and $\rho$ for
all~$x$  and finite values of~$t$. 
 It is clear that the structure of the 
 asymptotics must depend on the relation 
 between parameters $\varepsilon$ and $\rho$. 
We assume the fulfillment of the
condition 
 $$\mu=\displaystyle\frac{\rho}{\sqrt{\varepsilon}}\to 0.$$ 
 
 In paper~\cite{tmf}, a uniformly suitable asymptotic approximation 
 of the solution to problem $(\ref{eq})$--$(\ref{ic})$ 
 is constructed using the renormalization
method in the following most simple form. 
 Let us pass to the inner variables $$ 
 x = \sqrt{\varepsilon}\,\eta, 
 \qquad 
 t = \sqrt{\varepsilon}\,\theta, 
 $$ 
 since this allows one to take into account all
terms in equation (\ref{eq}). 
 As a first approximation, we take the solution
of 
 the equation 
 \begin{equation}\label{geq} 
 \frac{\partial Z}{\partial \theta} + 
 Z\frac{\partial Z}{\partial \eta} + 
 \frac{\partial^3 Z}{\partial \eta^3} = 0, 
 \end{equation} 
 with the initial condition 
 \begin{equation}\label{gic} 
 Z(\eta,0) = 
 \begin{cases} 
 \Lambda^-_0, & \eta<0, \\ 
 \Lambda^+_0, & \eta > 0, 
 \end{cases} 
 \end{equation} 
 where $\Lambda^\pm_0=\lim\limits_{s\to\pm\infty}\Lambda(s)$. 
 As a model of collisionless shock waves, 
 problem~(\ref{geq})--(\ref{gic}) 
 was studied by A.V. Gurevich and L.P. Pitaevskii 
 in~\cite{gp2}. 
 
 Let us construct the expansion of the solution in
the 
 following form: 
 \begin{equation}\label{rge} 
 u(x,t,\varepsilon,\rho) = 
 Z(\eta,\theta) + \mu W(\eta,\theta,\mu) + O(\mu^{\alpha}), 
 \qquad \alpha>0, 
 \end{equation} 
 where the addend $\mu W(\eta,\theta,\mu)$ must
eliminate 
 the singularity of $Z$ at the initial moment of
time. 
 Then the function~$W$ satisfies the
linear 
 equation 
 \begin{equation}\label{wle} 
 \frac{\partial W}{\partial \theta} + 
 \frac{\partial (Z W)}{\partial \eta} + 
 \frac{\partial^3 W}{\partial \eta^3} = 0. 
 \end{equation} 
 Differentiating equation~(\ref{geq}) 
 with respect to $\eta$, we find that 
 the expression $$ 
 G(\eta,\theta)= 
 \frac{1}{\Lambda_0^+ - \Lambda_0^-} 
 \frac{\partial Z(\eta,\theta)} 
 {\partial \eta} 
 $$ 
 satisfies equation~(\ref{wle}). 
 Moreover, $G$ is the Green function, 
 because $$ 
 \lim\limits_{\theta\to+0} 
 \int\limits_{-\infty}^{\infty} 
 G(\eta,\theta)f(\eta)\, d\eta = 
 - \frac{1}{\Lambda_0^+ - \Lambda_0^-} 
 \int\limits_{-\infty}^{\infty} 
 Z(\eta,0) f'(\eta)\, d\eta = f(0) 
 $$ 
 for any smooth function~$f$ with compact
support, 
 thus $G(\eta,0)=\delta (\eta)$.

 Let us choose the solution~$W$ in the
form 
 the convolution 
 with the Green function~$G$ 
 so that the asymptotic approximation would
satisfy 
 the initial condition~(\ref{ic}). 
 As a result, expansion~(\ref{rge})
becomes 
 $$ 
 u(x,t,\varepsilon,\rho) = U_0(x,t,\varepsilon,\rho) 
 + O(\mu^{\alpha}), 
 $$ 
 where 
 $$ 
 U_0(x,t,\varepsilon,\rho) = Z(\eta,\theta) + 
 \frac{\mu}{\Lambda^+_0 - \Lambda^-_0} 
 \int\limits_{-\infty}^{\infty} 
 \frac{\partial Z(\eta - \mu s,\theta)}{\partial \eta} 
 \left[ \Lambda(s) - Z(s,0)\right]\, ds. 
 $$ 
 Integrating by parts, we obtain the 
 asymptotic approximation 
$$
 u(x,t,\varepsilon,\rho) \approx 
 U_0(x,t,\varepsilon,\rho) = 
 \frac{1}{\Lambda^+_0 - \Lambda^-_0} 
 \int\limits_{-\infty}^{\infty} 
 Z\left( \frac{x-\rho s}{\sqrt{\varepsilon}}, \frac{t}{\sqrt{\varepsilon}}\right) 
 \Lambda'(s)\, ds. 
$$
 A rigorous mathematical justification of this formal
representation 
 is out of the framework of the present
paper. 
 It is clear that the behavior of the convolution
integral 
 is essentially determined by the solution
of~(\ref{geq})--(\ref{gic}) 
 with discontinuous initial data that
confirms 
 the importance of the Gurevich--Pitaevskii
problem. 
 
 Constructing complete asymptotic expansions 
 of the solution near the singular point 
 by the standard matching method 
 may be connected with serious difficulties. 
 In fact, it is necessary to solve 
 the scattering problem for a recurrence
system 
 of partial differential equations 
 with variable coefficients~\cite{ib}. 
 In addition, the investigation of the shock
wave 
 generated by gradient catastrophe shows
that 
 the asymptotics of the solution in a
neighborhood 
 of a singular point may have a multiscale
structure~\cite{vm4}. 
 
 The renormalization approach allows one
to 
 construct a uniformly suitable asymptotics 
 in the whole domain of independent
variables 
 avoiding difficulties arising 
 from the matching procedure. 

In particular,
 for $\Lambda^+_0=0$ and $\Lambda^-_0=a>0$
the following formula was obtained in paper~\cite{tmf}:
$$
u(x,t,\varepsilon,\rho) \approx
2\Lambda\left( \frac{x+at}{\rho}\right)
-\Lambda\left( \frac{x-2at/3}{\rho}\right)-
$$
$$
-\frac{at}{\rho}\int\limits_{-1}^{2/3}
\Lambda'\left( \frac{x- aty}{\rho}\right)
\left[2\, \mathrm{dn}^2
\left(\frac{a^{3/2} t\,\omega(y)}{\sqrt{\varepsilon}},
\sigma(y)\right) 
+\sigma^2(y)
\right] dy,
$$
where  $\mathrm{dn}(v,\sigma)$ is the elliptic Jacobi function
$$
\mathrm{dn}(u,m)= \sqrt{1-m\sin^2\varphi},
\qquad
u=\int\limits_{0}^{\varphi(u)}
\frac{dv}{\sqrt{1-m \sin^2 v}},
$$
$$
\omega(y)= \frac{1}{\sqrt{6}}
\left\{ y - \frac{1}{3}\left[ 1+\sigma^2(y)\right] \right\},
\qquad
1+\sigma^2-
\frac{2\sigma^2 (1-\sigma^2)K(\sigma)}
{E(\sigma)-(1-\sigma^2)K(\sigma)}=3y,
$$
$K(\sigma)$ and $E(\sigma)$ are
complete elliptic integrals of first and second kind:
$$
K(\sigma) = \int\limits_{0}^{\pi/2}
\frac{dv}{\sqrt{1-\sigma^2 \sin^2 v}}
\qquad
E(\sigma) = \int\limits_{0}^{\pi/2}
\sqrt{1-\sigma^2 \sin^2 v}\, dv.
$$
 
 Note that using the change $x=\rho\xi$, $t=\rho\tau$ 
 for~(\ref{eq})--(\ref{ic}), 
 we obtain the problem 
 $$ 
 \frac{\partial u}{\partial \tau} + 
 u\frac{\partial u}{\partial \xi} + 
 \mu^{-2} \frac{\partial^3 u}{\partial \xi^3} = 0, 
 \qquad 
 u\vert_{\tau=0} = \Lambda (\xi), 
 $$ 
 which formally contains only one parameter. 
 However, in this case one should study
asymptotics 
 as $\mu^{-2}\to\infty$ and $\tau\to\infty$. 
 
 \vskip 1cm 
\setcounter{equation}{0}
\section{Perturbation of a weak discontinuity}

Now consider the Cauchy problem for the KdV equation
\begin{equation}\label{kdv}  
 \frac{\partial u}{\partial t} + 
 u\frac{\partial u}{\partial x} +
 \frac{\partial^3 u}{\partial x^3} = 0
\end{equation}
with the initial condition
\begin{equation}\label{kic}
 u(x,0,\varepsilon) = \varepsilon \Lambda\left( \frac{x}{\varepsilon}\right), 
 \quad t=0,  \quad x\in\mathbb{R}, 
\end{equation}
where $\Lambda(s) = - s\, \Theta(-s) + O(s^{-2})$
as $s\to\infty,$
$\Theta$ is the Heaviside function.
By analogy with the case of the Burgers equation
we postulate the expression
\begin{equation}\label{ras}
R(x,t,\varepsilon) = \int\limits_{-\infty}^{\infty} 
 \frac{d^2\Lambda(s)}{ds^2} 
 \,\Phi(x-\varepsilon s, t)\, ds 
\end{equation} 
as an asymptotic solution for $0 < t\leqslant \delta < 1$, 
where $\Phi(x,t)$ is a smooth solution
of the Faminskii problem~\cite{famz}
\begin{equation}\label{fp}
 \Phi_t+\Phi\Phi_x+\Phi_{xxx}=0, 
\qquad \Phi(x,0)=-x\Theta(-x)
\end{equation} 
with an initial weak discontinuity
also called a contact discontinuity.
It is easy to show that the function (\ref{ras})
exactly satisfies the initial condition (\ref{kic}).

\vskip 5mm

\noindent
\textbf{Theorem~2.}
{\it Let $\Lambda(s)$ be a twice differentiable function and let
\begin{equation}\label{lb}
\Lambda(s) = - s\, \Theta(-s) + O(s^{-2}),
\qquad
\frac{d\Lambda(s)}{ds} = -\Theta(-s) + O(|s|^{-3}),
\qquad s\to\infty.
\end{equation}
Then the function $(\ref{ras})$
is an asymptotic solution of the Korteweg--de Vries equation $(\ref{kdv}).$
}
\vskip 2mm
\noindent
\textbf{Proof.} 
From (\ref{lb}) it follows that
$$
\int\limits_{-1/\sqrt{\varepsilon}}^{1/\sqrt{\varepsilon}}
\Lambda''(s) ds =  1+ O(\varepsilon^{3/2}),
\qquad
\int\limits_{-1/\sqrt{\varepsilon}}^{1/\sqrt{\varepsilon}}
s \Lambda''(s) ds = O(\varepsilon).
$$
Taking into account these relations, using (\ref{fp}),
and substituting $u=R(x,t,\varepsilon)$ into (\ref{kdv}), we find
$$
R_t + RR_x + R_{xxx} =
\int\limits_{-\infty}^{\infty} \Lambda''(s)
\Phi(x-\varepsilon s,t) \, ds
\int\limits_{-\infty}^{\infty} \Lambda''(s)
\Phi_x(x-\varepsilon s,t) \, ds -
$$
$$
- \int\limits_{-\infty}^{\infty} \Lambda''(s)
\Phi(x-\varepsilon s,t) \Phi_x(x-\varepsilon s,t) \, ds
 = O(\varepsilon^{3/2}).
$$

\bigskip
\textbf{Acknowledgments.}
This work was supported by the Russian Foundation for Basic Research,
project no.~14-01-00322.


\begin{thebibliography}{99}

\bibitem{bu}
J. Burgers, \textit{A Mathematical Model Illustrating the Theory of Turbulence},
 Advances in Applied Mechanics, Academic
Press, New York, 1948.

\bibitem{Gbw}
 G.B. Whitham, \textit{Linear and Non-Linear Waves}, Wiley-Interscience, 
New York, 1974. 

\bibitem{bk2}
 J. Bricmont, A. Kupiainen,
\textit{Renormalizing Partial Differential Equations},
 Lecture Notes in Physics, Springer-Verlag, 1994. 

\bibitem{gv}
 N. Goldenfeld, J. Veysey,
\textit{Rev. Mod. Phys.}, 79 (2007), 883-927. 

\bibitem{teodor}
 E.V. Teodorovich,
\textit{J. Appl. Math. Mech.}, 68:2 (2004), 299-326. 

\bibitem{bkl}
J. Bricmont, A.Kupiainen, G. Lin, 
\textit{Comm. Pure Appl. Math.}, 47 (1994), 893-922. 

\bibitem{zvm}
 S.V. Zakharov,
\textit{Comp. Math. Math. Phys.}, 50:4 (2010), 665-672. 

\bibitem{vm4}
 S.V. Zakharov,
\textit{Comp. Math. Math. Phys.}, 44:3 (2004), 506-513. 

\bibitem{gst}
 R. Garifullin, B. Suleimanov, N. Tarkhanov,
\textit{Phys. Lett. A}, 374:13 (2010), 1420-1424. 

\bibitem{kf}
{T.~Kappeler},
\textit{J. Diff. Eq.}
63:3, (1986), 306-331.

\bibitem{fam}
A.V. Faminskii, 
\textit{Math. USSR-Sb.} 68:1 (1991), 31-59. 

\bibitem{tmf}
 S.V. Zakharov,
\textit{Theor. Math. Phys.}, 175:2 (2013), 592-595.

\bibitem{gp2}
 A.V. Gurevich, L.P. Pitaevskii,
\textit{Sov. Phys.-JETP}, 38:2 (1974), 291-297. 

\bibitem{ib}
 A.M. Il'in,
\textit{Matching of Asymptotic Expansions of Solutions of Boundary Value Problems},
AMS, Providence, RI, 1992. 

\bibitem{famz}
A.V. Faminskii,
\textit{Math. Notes}, 83:1 (2008), 107-115.


\end{thebibliography}
\end{document}